\documentclass[english]{article}
\pdfoutput=1

\usepackage[T1]{fontenc}
\usepackage{geometry}
\usepackage{setspace}
\usepackage[latin9]{inputenc}
\usepackage{color}
\usepackage{babel}
\usepackage{verbatim}
\usepackage{float}
\usepackage{amsmath}
\usepackage{amsthm}
\usepackage{amssymb}
\usepackage{graphicx}
\usepackage{esint}
\usepackage[colorlinks=true,pdftitle={Trading styles and long-run variance of asset prices}]{hyperref}
\usepackage[authoryear]{natbib}
\usepackage{bm}

\floatstyle{ruled}
\newfloat{algorithm}{tbp}{loa}
\providecommand{\algorithmname}{Algorithm}
\floatname{algorithm}{\protect\algorithmname}

\theoremstyle{plain}

\theoremstyle{plain}
\newtheorem{thm}{\protect\theoremname}
\theoremstyle{plain}
\newtheorem{prop}{\protect\propositionname}
\theoremstyle{plain}

\usepackage{dsfont}

\usepackage{subfig}
\makeatother

\providecommand{\assumptionname}{Assumption}
\providecommand{\lemmaname}{Lemma}
\providecommand{\propositionname}{Proposition}
\providecommand{\theoremname}{Theorem}

\usepackage{adjustbox} 

\begin{document}
\title{Trading styles and long-run variance of asset prices}
\author{Lawrence Middleton\thanks{Research sponsored by The Witness Corporation (www.thewitnesscorporation.com)},
James Dodd\footnotemark[1], Simone Rijavec\footnotemark[1]\ \footnote{University of Oxford, Department of Physics}}
\maketitle
\begin{abstract}
Trading styles can be classified into either trend-following or mean-reverting.
If the net trading style is trend-following the traded asset is more
likely to move in the same direction it moved previously (the opposite
is true if the net style is mean-reverting). The result of this is
to introduce positive (or negative) correlations into the time series.
We here explore the effect of these  correlations on the long-run
variance of the series through probabilistic models designed to explicitly
capture the direction of trading. Our theoretical insights suggests
that relative to random walk models of asset prices the long-run variance
is increased under trend-following strategies and can actually be
\emph{reduced} under mean-reversal conditions. We apply these models
to some of the largest US stocks by market capitalisation as well
as high-frequency EUR/USD data and show that in both these settings,
the ability to predict the asset price is generally \emph{increased}
relative to a random walk. 
\end{abstract}

\section{Background and related literature}

The variance of a financial asset is a proxy for how predictable the
asset is after some given period. The following aims to quantify the
change in variance of an instrument under the simplifying asssumption
that the price dynamics are governed by either prevailing trend-following
strategies or prevailing mean-reverting strategies. On the one hand,
if the overall view of the market is momentum based then an increase
in the price on any trading day is likely to increase the share price
on the following day \citep{chan1996momentum,jegadeesh2001profitability}.
Conversely, if the overall view of the market is mean reverting then
an increase is likely to be followed by a successive decrease the
following day. Importantly, the former mimics the typical scenario
arising in passive investment, where for example additional positive
investment follows initially favourable conditions thereby driving
up the price of the asset. In the following, we investigate how each
of these styles impacts the overall variance of an instrument. 

We explore this probabilistically through examining a \emph{correlated
random walk} \citep{renshaw1981correlated} - a model such that the
returns take values either -1 or +1 and where a day's closing returns
is equal to the previous day's with probability $p$. Such a model
touches on random walk theory for asset pricing. There exists a long
history of using random walks in finance and econophysics \citep{fama1995random,sewell2011history,scalas2006five},
a full review is beyond the scope of this article. Of note also, random
walk models appear in the evaluation of the Efficient Market Hypothesis
\citep{malkiel2003efficient,malkiel1970efficient} where the predictability
of a process is central to establishing whether markets are efficient
or not. 

In addition to the simplistic model, we consider two models applicable
to financial assets over different return horizons. We consider firstly
daily US stock data where after developing and validating an appropriate
model we estimate the probability of each stock moving in the same
direction on subsequent days. Exploiting this, we are able to quantify
the change in variance relative to a random walk for each of the assets
and thereby show that the variance is \emph{deflated} in most of these
stocks. We also test the prediction accuracy of a similar model on
a week's worth of high-frequency EUR/USD where we are able to conclude
again that the net trading style is mean-reverting and to a more exagerated
extent than the daily stock data. 

\section{Simulation study}

We consider two discretet-time stochastic process models for asset
prices $(X_{t})$ and $(Y_{t})$ for $t=0,1,...$. The first, (RW)
is a random walk where $\mathds{P}[X_{t+1}=X_{t}+1]=\mathds{P}[X_{t+1}=X_{t}-1]=\frac{1}{2}$.
The second,  random walk (CRW) is a random walk with persistence in
the direction it travels, i.e., $Y_{t}$ for $t=0,1,...$
\begin{enumerate}
\item If $Y_{t}=Y_{t-1}+1$ (last move was up) then $Y_{t+1}=Y_{t}+1$ (next
move will be up) with probability $p$
\item If $Y_{t}=Y_{t-1}-1$ (last move was down) then $Y_{t+1}=Y_{t-1}-1$
(next move will be down) with probability $p$
\end{enumerate}
\begin{figure}
 \begin{adjustbox}{center}\subfloat[Random walk with persistence $(Y_{t})$ $p=\frac{1}{4}$]{\centering{}\includegraphics[scale=0.6]{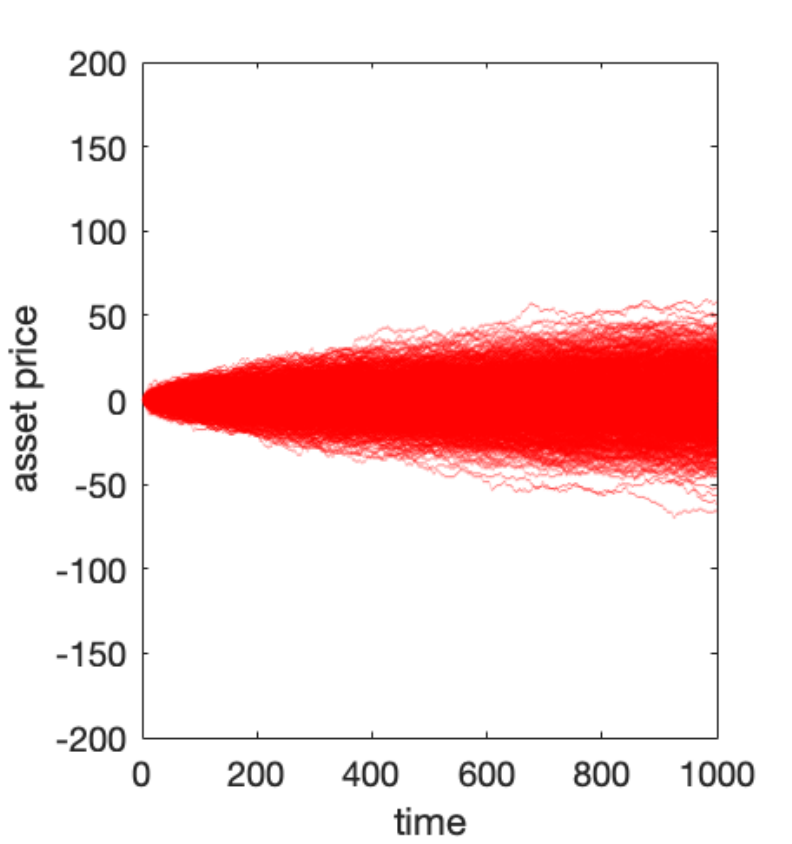}}\subfloat[Random walk $(X_{t})$]{\centering{}\includegraphics[scale=0.6]{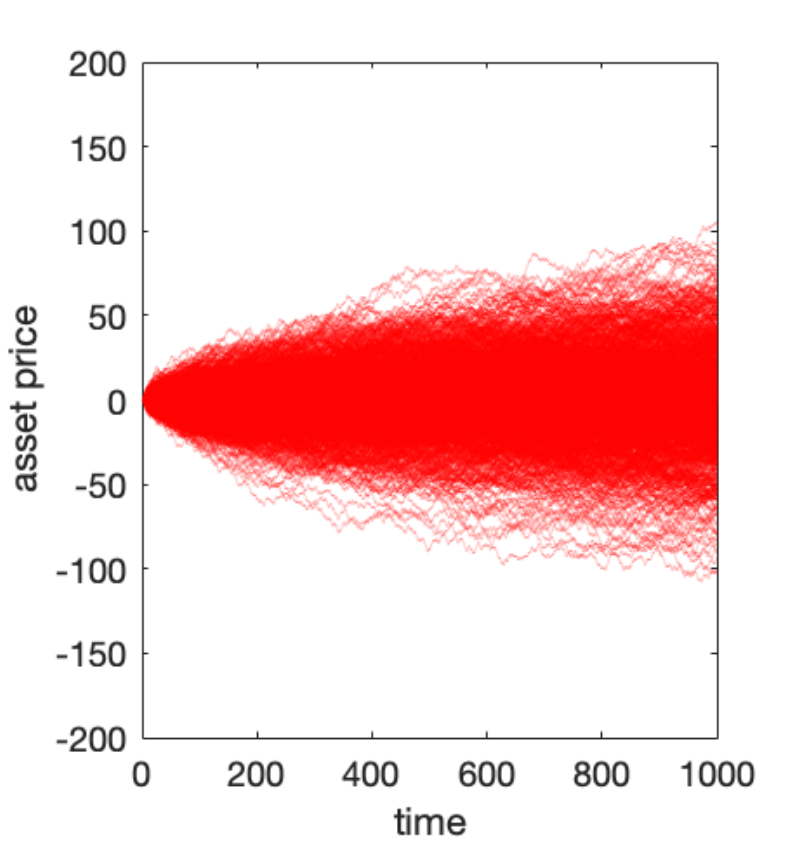}}\subfloat[Random walk with persistence $(Y_{t})$ $p=\frac{3}{4}$]{\centering{}\includegraphics[scale=0.6]{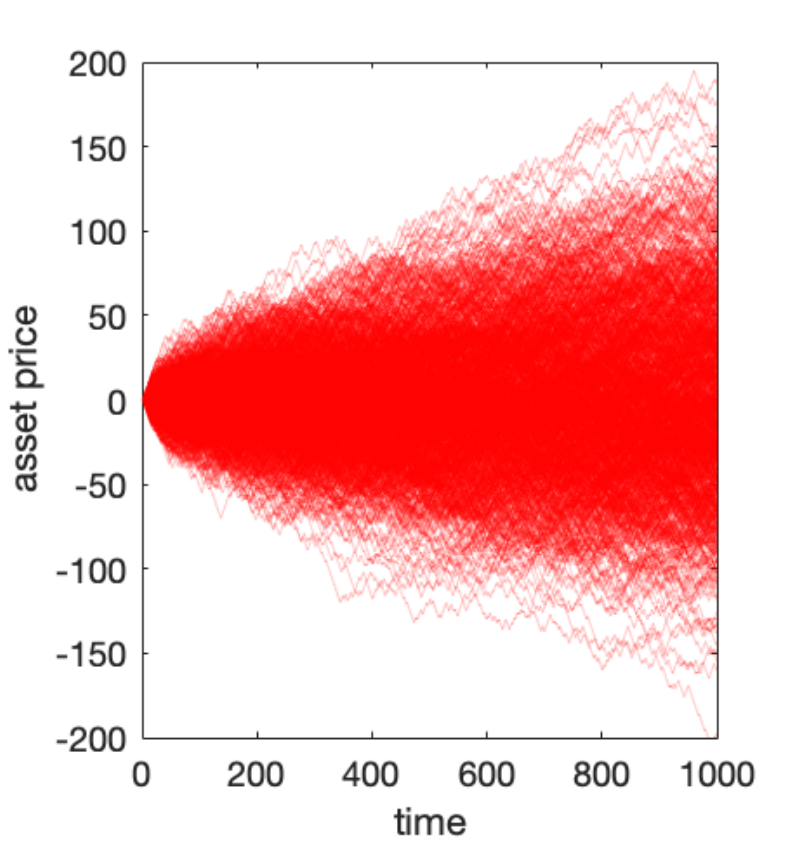}}
\centering{} \end{adjustbox} \caption{Simple models for asset prices}
\label{fig:assetpricerealisations}
\end{figure}

\begin{figure}
\centering{} \begin{adjustbox}{center}\subfloat[Comparison of standard deviation through time for RW and CRW.]{\begin{centering}
\includegraphics[scale=0.6]{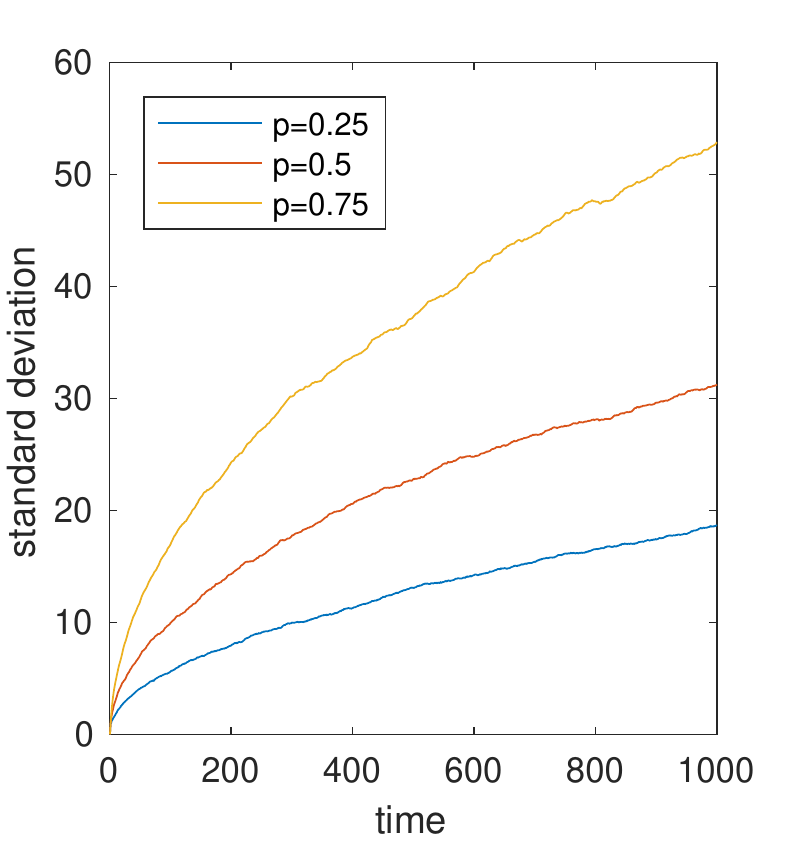}
\par\end{centering}
\label{fig:relstd}}$\qquad$\subfloat[Comparison variance for RW and CRW.]{\begin{centering}
\includegraphics[scale=0.6]{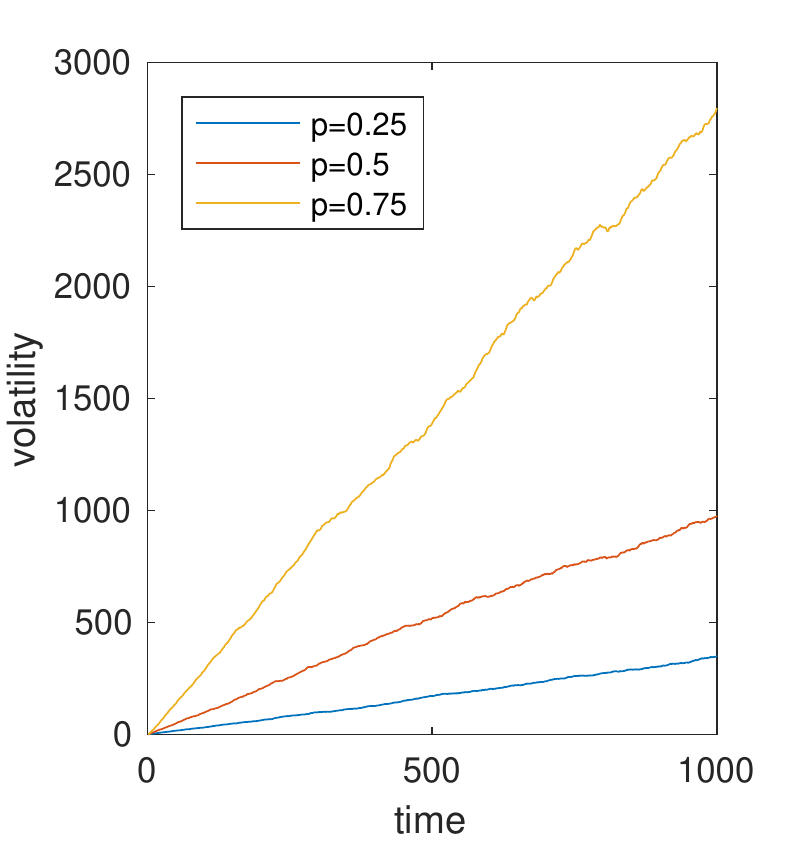}
\par\end{centering}
\label{fig:relvol}}$\qquad$\subfloat[Change in variance at $T=200$ as a function of $p$ - normalised
by the variance at $p=\frac{1}{2}$.]{\begin{centering}
\includegraphics[scale=0.65]{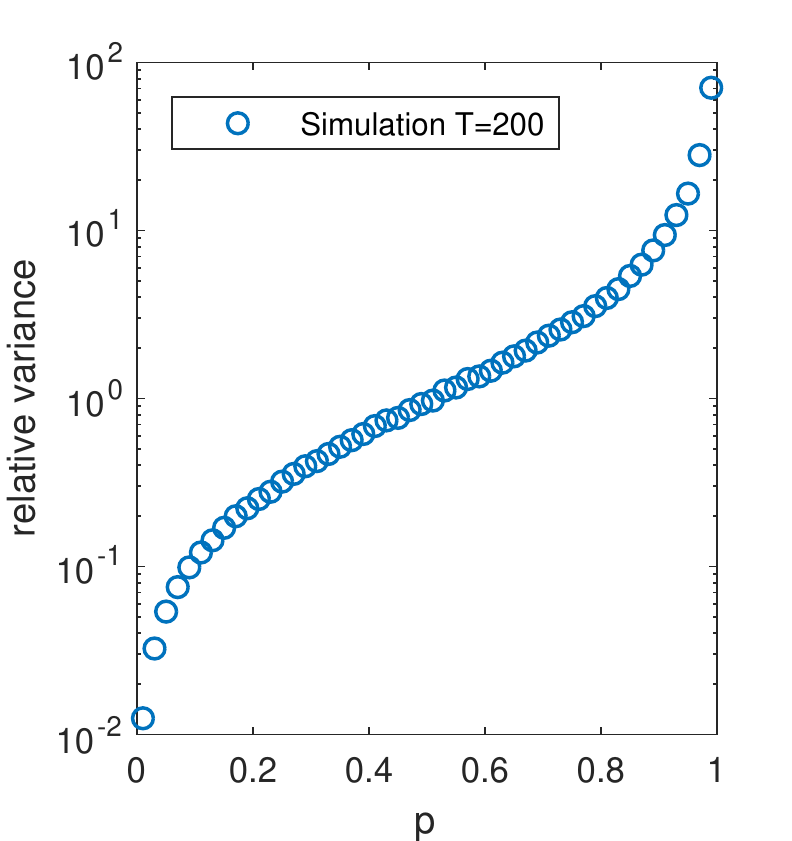}
\par\end{centering}
 \label{fig:relvolra}} \end{adjustbox} \caption{Comparison of measures of uncertainty for two models.}
\end{figure}

We plot these two situations in Figure \ref{fig:assetpricerealisations},
simulating 1000 realisations of each of the two models over 1,000
units of time, setting $p=\frac{1}{4}$ and $p=\frac{3}{4}$ for CRW
and where the initial directions were distributed randomly up and
down with probability 1/2. We see that there appears to be a relative
increase in the variance of the assets for the random walk with persistence
compared to that without when $p=\frac{3}{4}$ and a relative decrease
when $p=\frac{1}{4}$. As a result, we look at the standard deviation
of each of the two models through time in Figure \ref{fig:relstd}
where we see the standard deviation is close to double that of the
random walk model when $p=\frac{3}{4}$. Indeed, in Figure \ref{fig:relvol}
the variance (square of std. deviation.) is approximately linear in
both cases.

Next, we examine the relative increase in variance between the two
models as we vary $p$, the probability we move in the same direction
as the previous day. We show $\mathds{V}[Y_{T}]/\mathds{V}[X_{T}]$
in Figure \ref{fig:relvolra} where $T=200$ is the final day. Interestingly,
we see that over a broad range of $p$, e.g. $p\in[0.2,0.8]$ then
the relative increase in variance is approximately linear in the log-scale
(suggesting it is exponential in $p$) and for $p$ close to $1$
the relative increase in volatiliy rises super-exponetially. We note,
that the variance reduces for $p<0.5$ as in this case the model of
CRW if the last move was down (for example) the next move will be
up with high probability and so as $p\rightarrow0$ then the asset
price tends to something that alternates between $0$ and $+1$ or
0 and $-1$. Finally, assuming that in both cases the variance is
approximately linear in time, we are able to take the ratio of the
volatilities between the two to give some measure of the relative
change in variance per unit of time as well.

\section{Application to daily data}

We here introduce a semiparametric model for a price process, relaxing
the assumption that the increments must be in $\{-1,+1\}$ to a more
general class of return distributions. We consider discretely observed
price processes, with consideration of continuous-time processes to
follow. 

We assume, as before, that the sign of the increment is governed by
a correlated random walk model, with increments taking values in $\{-1,1\}$,
which we denote as $(S_{n})$. Conditional on $S_{n-1}$, then $S_{n}=S_{n-1}$
with probability $p$ and $S_{n}=-S_{n-1}$ with probability $1-p$.
Coupled with this, we also introduce random variables representing
the magnitude of the change in price at each iteration, which we denote
as $(\delta_{n})$. The overall price process may then be expressed
as 
\begin{align}
Z_{n} & =\sum_{i=0}^{n}S_{i}\delta_{i}\label{eq:semipflex}
\end{align}
where $\delta_{i}\ge0$ are random variables used to model an increment
of an arbitrary price process. We assume that conditional on a realisation
of $(S_{i})_{i=0}^{n}$ the increments are conditionally independent,
with $\delta_{i}\stackrel{i.i.d.}{\sim}\nu$ for some distribution
$\nu$.
\begin{prop}
\label{prop:semipprop}For $S_{0}$ symmetrically distributed on $\{-1,1\}$
and $\mathds{E}_{\nu}[\delta^{2}]<\infty$, the limiting variance
of the proposed price process satisfies
\begin{equation}
\lim_{n\rightarrow\infty}\frac{1}{n}\mathds{V}[Y_{n}]=\mathds{E}_{\nu}[\delta^{2}]+\mathds{E}_{\nu}[\delta]^{2}\left(\frac{2p-1}{1-p}\right)\label{eq:varsemip}
\end{equation}
where $\mathds{E}_{\nu}$ denotes the expectation of a random variable
distributed according to $\nu$. 
\end{prop}

Proof is in Appendix \ref{subsec:proofprop}. Previous expressions
for limiting variance of a correlated random walk model can be found
in \citep{renshaw1981correlated,mauldin1996directionally,guo2017optimal},
though without the additional complexity arising from a random magnitude
distribution $\nu$. 

\begin{figure}
\begin{centering}
\includegraphics[scale=0.55]{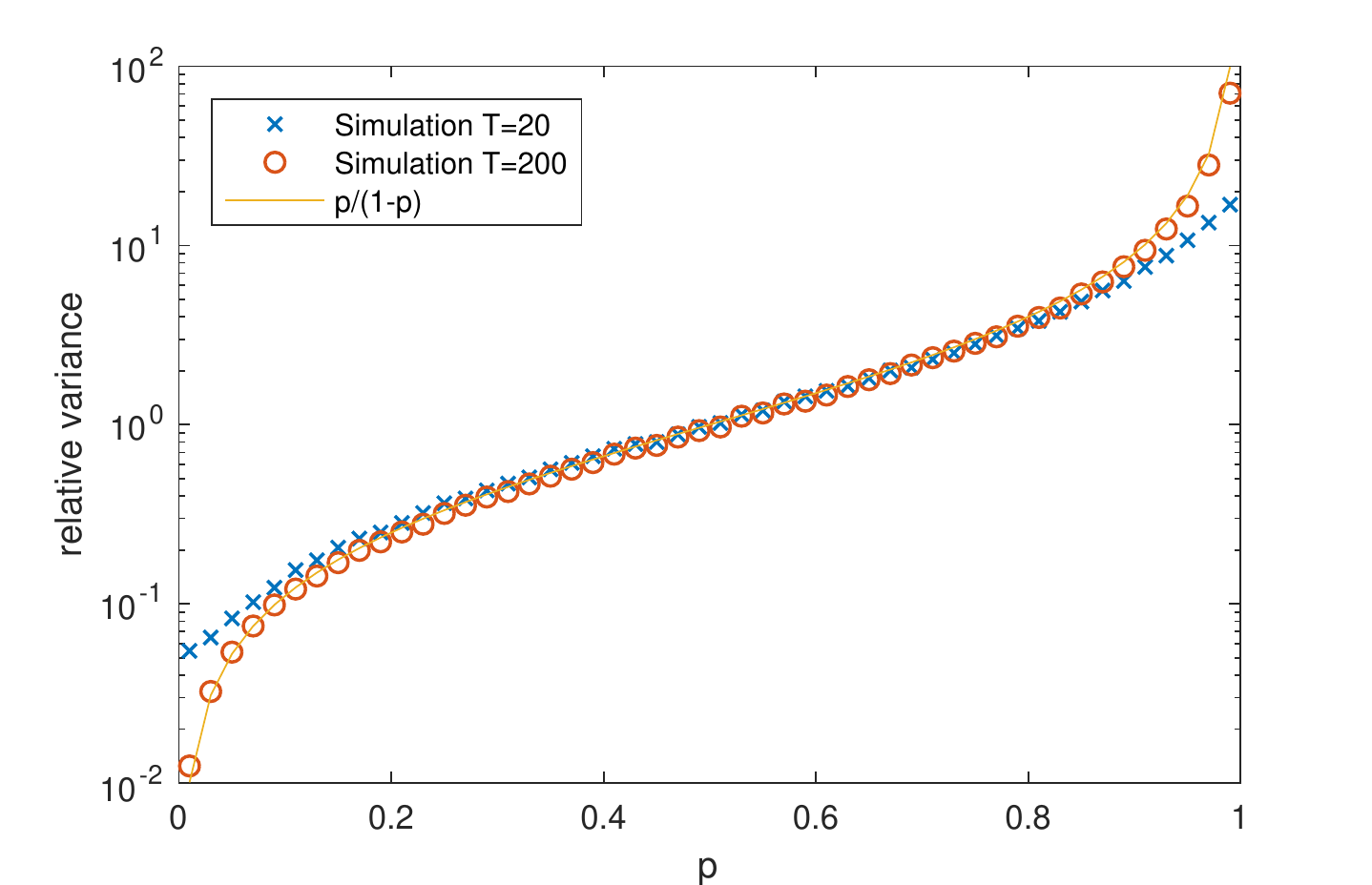}
\par\end{centering}
\caption{Change in variance at $T=200$ as a function of $p$. Shown also is
the change in variance as a function of $p$ for $T=20$ where we
see that the change the solution $p/1-p$ breaks down in the tails
though captures the central part well.}
\label{fig:relvol_exp}
\end{figure}

Considering Equation (\ref{eq:varsemip}), we see that for a magnitude
distribution $\nu$ concentrated on a single (strictly positive) point
$x$, then the $\mathds{E}_{\nu}[\delta^{2}]=\mathds{E}_{\nu}[\delta]^{2}$
and the limiting variance is 
\begin{equation}
\lim_{n\rightarrow\infty}\frac{1}{n}\mathds{V}[Y_{n}]\propto\frac{p}{1-p}\label{eq:p1minusp}
\end{equation}
thereby capturing the previous analysis as a special case. Indeed,
we see that in this case for $p>\frac{1}{2}$ then the the variance
is increased and for $p<\frac{1}{2}$ the reverse is true, indeed
the  variance is a montonically increasing function of $p$. We compare
the change in variance as a function of $p$ with the previous simulation
estimates, overlaying the analytical solution in Figure \ref{fig:relvol_exp}.
We see that $p/1-p$ fits the experiment exceptionally well for $T=200$.
Additionally we see that for $T=20$ the fit is less good in the extremal
values of $p$ though still permissible. Such a result suggests that
the result is indeed asymptotic (i.e. the change in variance is not
necessarily $p/1-p$ over a small period) though in any case we expect
the long run behaviour to better capture the current dynamics of an
asset price. Finally, we also see according to Proposition \ref{prop:semipprop}
that a random walk model can be recovered through setting $p=\frac{1}{2}$,
in which case $\lim_{n\rightarrow\infty}\frac{1}{n}\mathds{V}[Y_{n}]=\mathds{E}_{\nu}[\delta^{2}]$. 

\subsection{Model estimation}

The model proposed in (\ref{eq:semipflex}) depends on a parameter
$p$, modelling the distribution of the sign process, and the magnitude
of the returns $(\delta_{n})_{n\ge0}$. We see that the sequence $(S_{n})_{n\ge0}$
is an observable Markov chain parameterised by $p$ - where $p$ can
be estimated through the proportions of consecutive $S_{n}$ that
share the same sign. The following will make no further assumptions
on the distribution of $\nu$, the magnitude distribution, beyond
the previously mentioned condition that $\mathds{E}_{\nu}[\delta^{2}]<\infty$.

\subsection{US stock data}

\begin{figure}[H]
\begin{centering}
\includegraphics[scale=0.5]{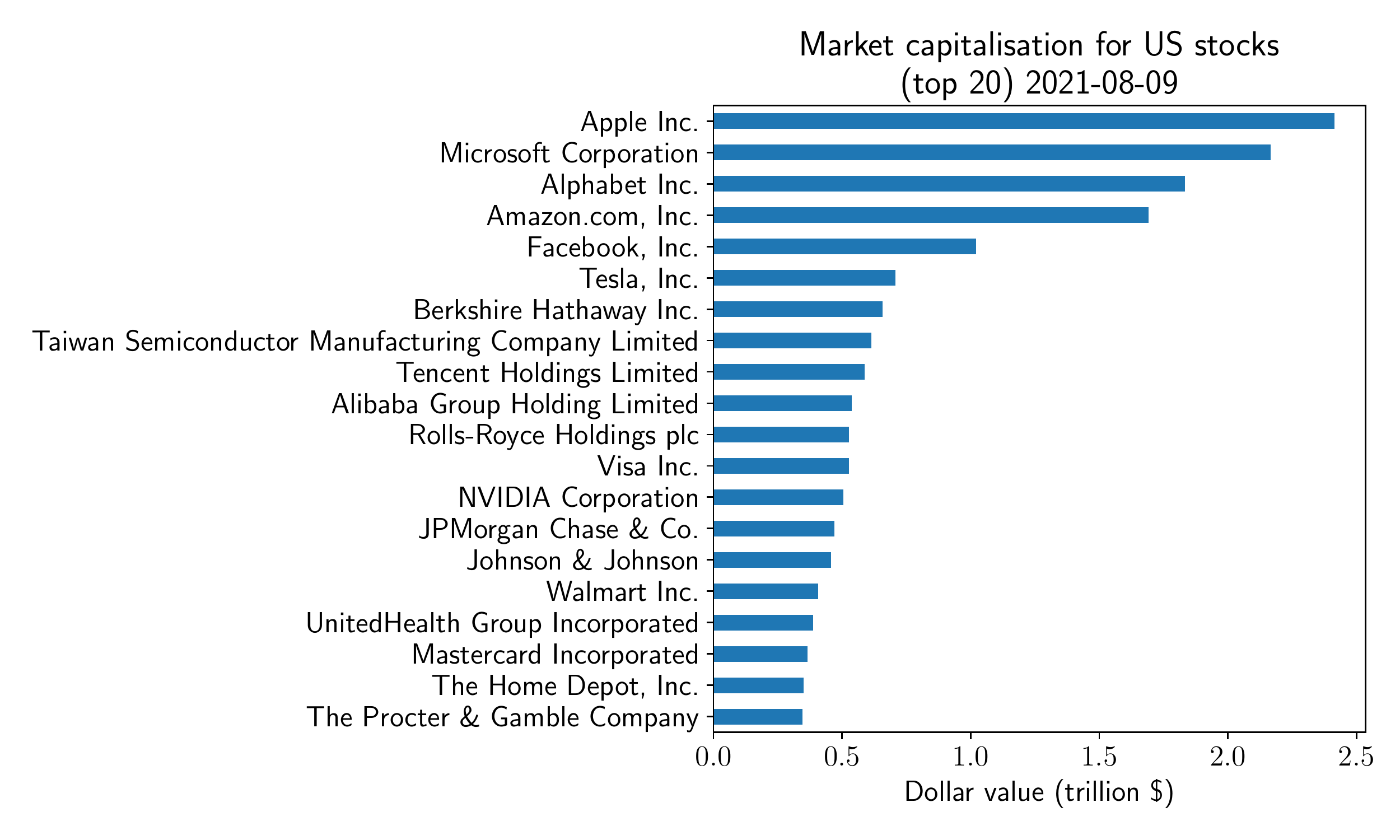}
\par\end{centering}
\caption{Market capitalisation for largest 20 stocks.}
\label{fig:marketcap}

\end{figure}

We exploit Proposition \ref{prop:semipprop} to find the relative
increase in variance of asset prices arising from persistence in trading
directions. We focus primarily on the US stock market, in particular
focussing on some of the largest stocks available for trading. We
restrict attention to those stocks with traded volume greater than
or equal to 100,000 and extract daily closing data for the 100 largest
stocks (sorted by market capitalisation). A complete list of these
stocks is available in the Appendix \ref{subsec:usstocktables} and
the market cap. for the largest 20 is shown in Figure \ref{fig:marketcap}. 

The time period considered was three years of daily data from $9^{th}$
August 2018 to $9^{th}$ August 2021, with data extracted from Yahoo
Finance, using \texttt{yfinance} Python package. A modest amount of
filtering was performed to ensure only suitable stocks were included
in the analysis. In particular, two stocks were excluded based on
large observed periods of no change in the price. Finally, Alphabet
Inc. was duplicated with GOOGL and GOOG ticker symbols (stocks with
and without voting rights) so only the largest of the two was included
(GOOGL). This resulted in a total of 97 of the 100 stocks being included,
a complete list of these can be found in the Appendix.

Crucial to the subsequent analysis is some guarantee that the assumptions
made in Proposition \ref{prop:semipprop} are reasonable. As such,
we perform two model validations, checking firstly that the distribution
of returns is approximately symmetric and secondly that the magnitude
of the increments is uncorrelated with the sign of the increments.
Details around these two procedures are provided in the Appendix \ref{subsec:USmodelvalidation},
though in summary, at a significance level of 5\%, approximately 5.15\%
of the stocks included in the analysis did not show sufficient evidence
of either asymmetric increments or increment magnitude correlated
with sign.

Having performed this model validation, it is then possible to estimate
the increase in variance arising from persistence in trading direction
through estimating quantities appearing in Equation (\ref{eq:varsemip}).
In particular we are interested in the change in variance caused by
persistent market trading. We denote this $\bar{\sigma}^{2}$ and
express it as 
\[
\bar{\sigma}^{2}:=1+\frac{\mathds{E}_{\nu}[\delta]^{2}}{\mathds{E}_{\nu}[\delta^{2}]}\left(\frac{2p-1}{1-p}\right)
\]
Estimates of $\mathds{E}_{\nu}[\delta^{2}]$ and $\mathds{E}_{\nu}[\delta]$
are obtained through sample averages of the distribution of magnitudes
of the price process. Return increments were taken as additive on
the log-scale, so that $\delta_{i}=|\log(Z_{i})-\log(Z_{i-1})|$.
Based on estimates of $\mathds{E}_{\nu}[\delta^{2}]$, $\mathds{E}_{\nu}[\delta]$
and $p$ from historical data we are able to compare the inflated
variance due to persistence with the natural variance arising from
a random walk model.

\begin{figure}
 \begin{adjustbox}{center}\subfloat[Distribution of estimates of $p$ for US stocks]{\begin{centering}
\includegraphics[scale=0.45]{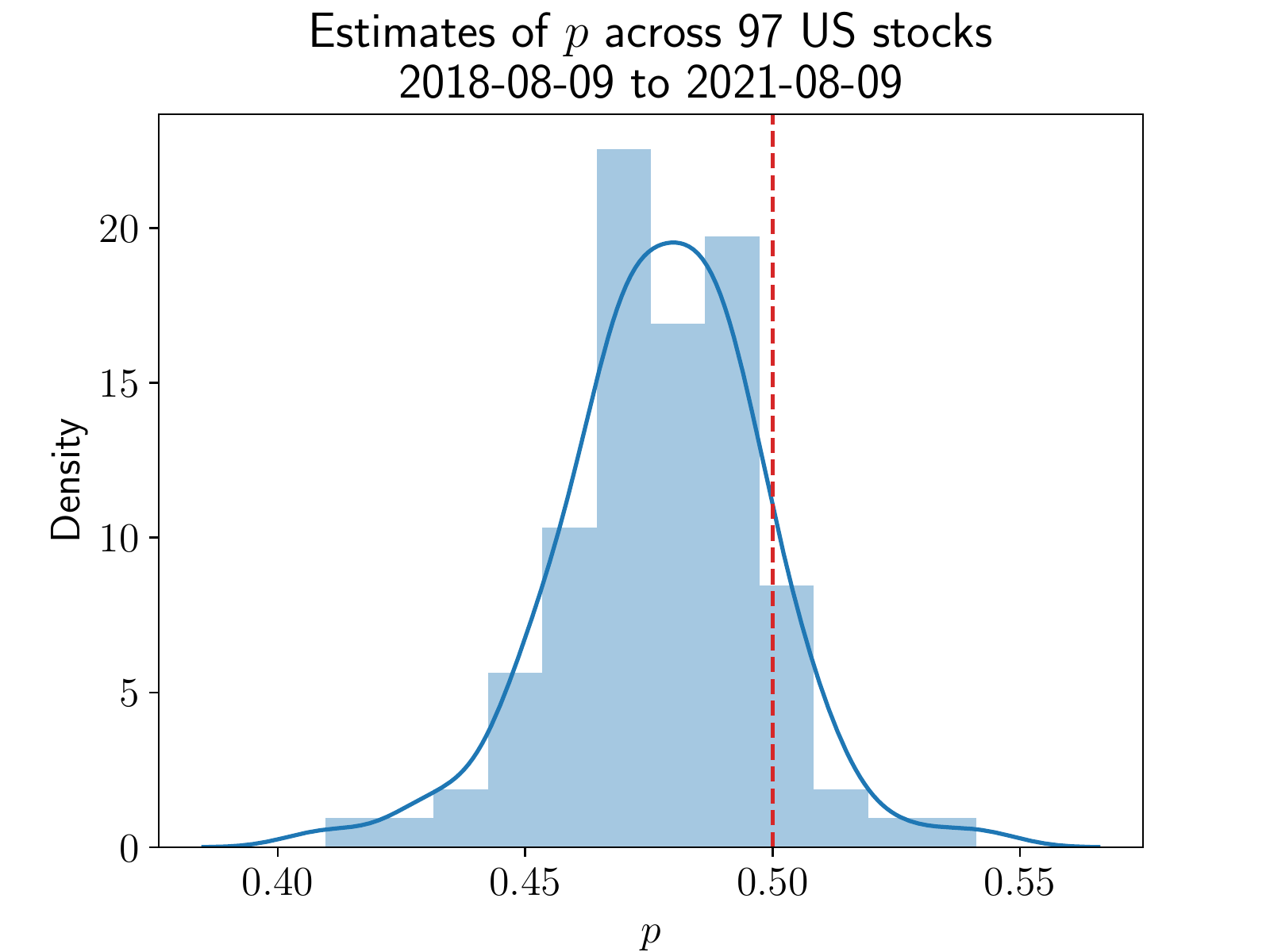}
\par\end{centering}
\label{fig:pUS}}\subfloat[Distribution of estimates of $\bar{\sigma}^{2}$ across US stocks]{\begin{centering}
\includegraphics[scale=0.45]{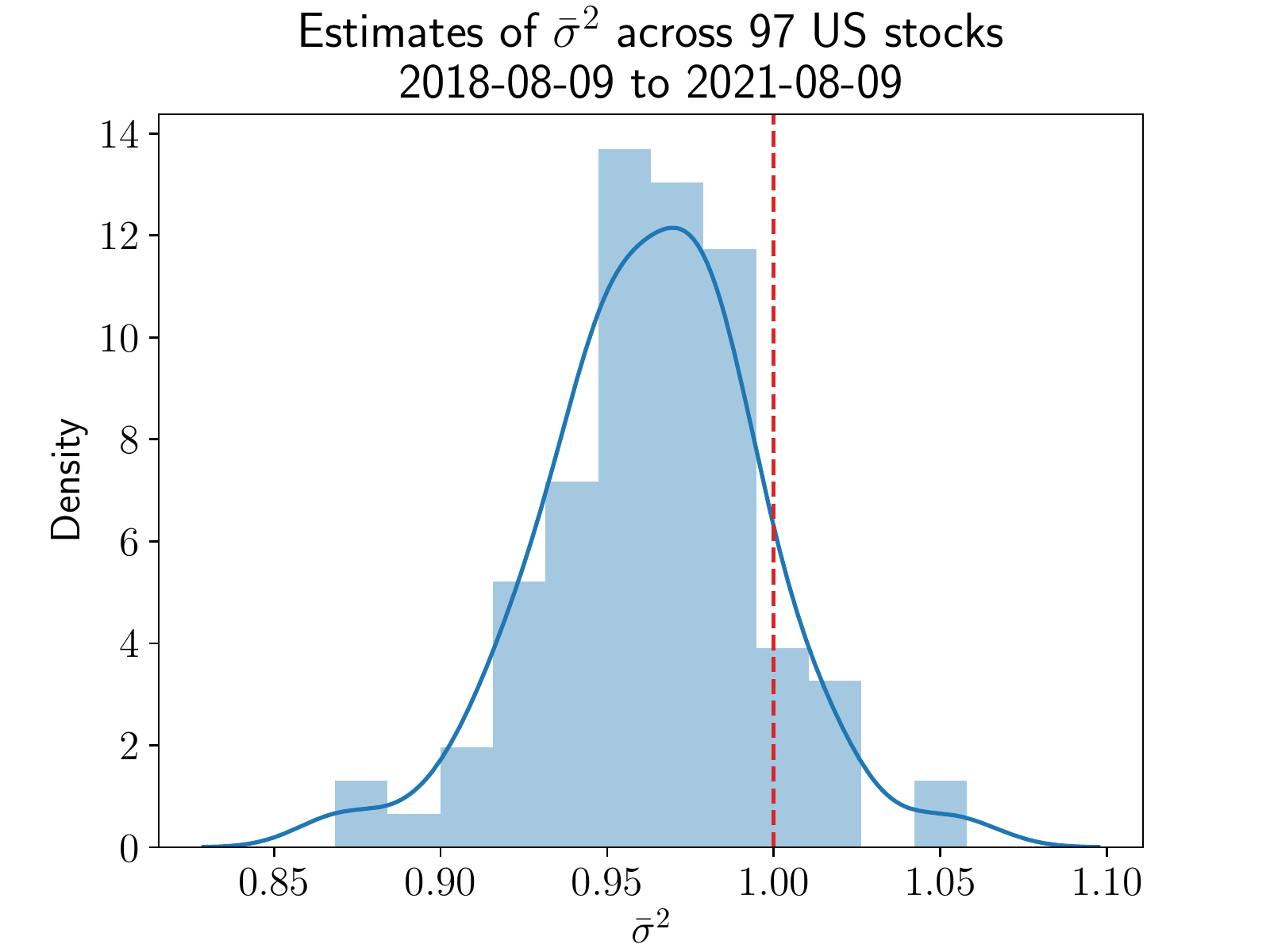}
\par\end{centering}
\label{fig:relvarUS}} \end{adjustbox} 
\centering{}\caption{Inflation of variance of US stocks relative to random walk}
\end{figure}

In Figure \ref{fig:pUS} we show the estimated values of $p$ across
the 97 stocks under consideration. Adjacent to this we show the distribution
of variances, normalised by the variance of a random walk model. Strikingly
we see that typical values of $p$ are less than $\frac{1}{2}$ and
the variance of the assets are actually \emph{deflated}. Such a result
suggests that in some of the world's largest stocks, the freneticism
apparent in the market actually serves to make the resulting processes
\emph{more }predictable - their variance is lower. Specifically we
see that the median value of $\bar{\sigma}^{2}$ is 0.965, with lower
and upper quartile given by 0.946 and 0.982 respectively. In summary,
we see that there is approximately a 1.7-5.4\% \emph{reduction} in
variance due to persistence in trading directions.

\section{Application to high-frequency data}

In addition to the discrete-time analysis in the previous section
we also develop methodology capable of learning $p$ in a flexible
model of high-frequency data. Such a model may also be applicable
to highly illiquid assets, exhibiting only occasional changes in price. 

As a result, rather than considering trades occuring at regular intervals
we assume the times at which a price changes is itself a stochastic
process. The resulting model falls under the class of Markov renewal
processes \citep{pyke1961markov}. They are closely related to the
continuous time random walks of \citep{kotulski1995asymptotic,tunaley1974asymptotic,scalas2006application}.
In the physics literature continuous time random walks have found
applications in econophysics and finance, see \citep{scalas2006application,scalas2006five}
for some summarising results illustrating the deep connectiom between
continuous time random walks and fractional calculus. We show in Appendix
\ref{subsec:continuoustimeanalysis} that much of the previous analysis
in discrete time generalises to this setting under the critical assumption
that the arrival times between events are independent of the sign
and magnitude process.

\subsection{A high-frequency model}

As remarked in \citep{scalas2006application}, the distribution increments
of event times is typically coupled to the distribution of jumps.
The following proposes a general model motivates by the previous analysis
to allow for a more flexible distribution over inter-event times and
jump values. Importantly, despite the additional sophistication, the
model remains interpretable and retains a probability parameter $p$
of moving in the same direction as the previous iteration. 

To better capture the dynamics at higher frequencies, where the time
between events is itself a random variable, we model jointly the price
process and the time-to-event process. Let $Z_{t}$ denote the state
of the process at time $t$. Then with $\tau_{n}$ denoting the time
between events $n-1$ and $n$ , and $(S_{n},\delta_{n})$ denoting
the sign and magnitude process defined previously, then the process
at iteration $n$ evolves as follows
\begin{align}
S_{n}|\{S_{n-1}=s_{n-1}\} & \sim P(\cdot|s_{n-1})\label{eq:sophm1}\\
(\delta_{n},\tau_{n})|\{S_{n}=s_{n},S_{n-1}=s_{n-1}\} & \sim f(\cdot|s_{n},s_{n-1})\label{eq:sophm2}
\end{align}
where $P(\cdot|s)$ denotes the Markov transition kernel corresponding
to the previous defined Markov sign process. The additional distribution
$f$ specifies the joint distributions over the increments of the
price process and the increments of time-to-event process. The final
price process is then given as
\[
Z_{t}=\sum_{i=1}^{N(t)}S_{i}\delta_{i}\quad\text{where}\quad N(t)=\sup\{n:\sum_{i=1}^{n}\tau_{i}\le t\}
\]
with the latter denoting the counting process.

Analogously to the simplified model, using $p$ we are able to control
the direction of the price process (either positive or negative) with
sizes of the increment specified by the more flexible distribution
$f$.

Where possible, we opt to use the empirical distribution of historical
increments, in particular, to define the choice of $f$. Based on
this model, we are able to construct a flexible model capable of simulating
from (\ref{eq:sophm1}) and (\ref{eq:sophm2}) through selecting $f$
to be the empirical distribution of $(\delta_{n},\tau_{n})$ conditioned
such that 
\begin{align}
(\delta_{n},\tau_{n})|(s_{n},s_{n-1}) & \sim\begin{cases}
\hat{F}(\{\delta_{n},\tau_{n}:\delta_{n}>0,\delta_{n-1}>0\}) & if\quad s_{n}=1,s_{n-1}=1\\
\hat{F}(\{\delta_{n},\tau_{n}:\delta_{n}>0,\delta_{n-1}\leq0\}) & if\quad s_{n}=1,s_{n-1}=-1\\
\hat{F}(\{\delta_{n},\tau_{n}:\delta_{n}\leq0,\delta_{n-1}>0\}) & if\quad s_{n}=-1,s_{n-1}=1\\
\hat{F}(\{\delta_{n},\tau_{n}:\delta_{n}\leq0,\delta_{n-1}\leq0\}) & if\quad s_{n}=-1,s_{n-1}=-1
\end{cases}\label{eq:empdists}
\end{align}
where we use $\hat{F}(\{x_{i}\})$ denote the empirical distribution
function of datapoints $\{x_{i}\}$.

\subsubsection{Forecasting high-frequency EUR/USD}

We fit the model to two weeks of high-frequency EUR/USD commencing
$8^{th}$ March 2021, where week 1 was used as data to estimate the
empirical distribution in (\ref{eq:empdists}) and where week 2 was
used to assess the quality of wthe forecasts. The high-frequency data
was provided by the Dukascopy exchange and partially discretised,
so that the minimum time between events was 0.05s. 

\begin{figure}[H]
\begin{centering}
 \begin{adjustbox}{center}\subfloat[Forecasts \& historical data $p=\frac{1}{4}$]{\begin{centering}
\includegraphics[scale=0.5]{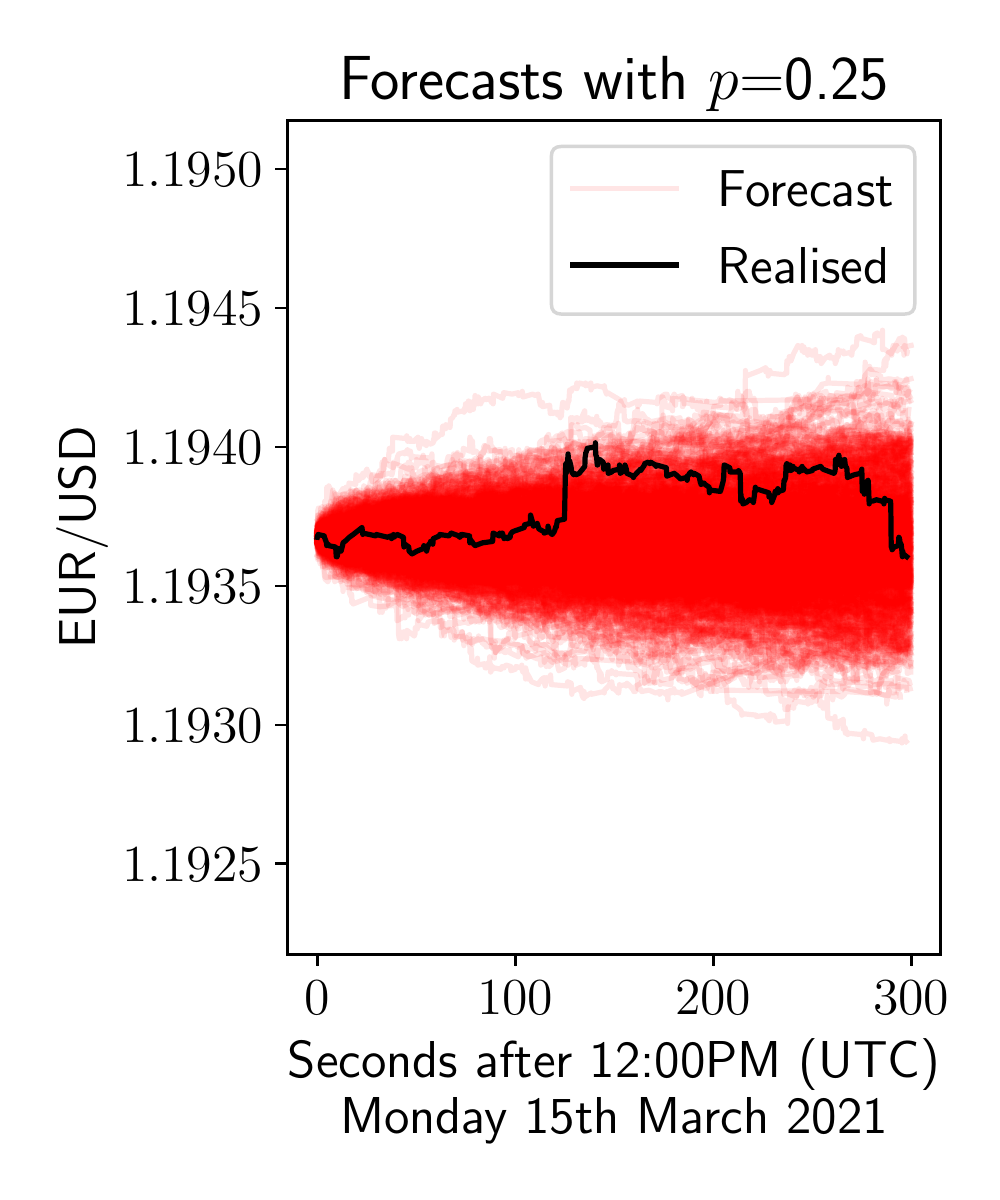}
\par\end{centering}
}\subfloat[Forecasts \& historical data $p=\frac{1}{2}$]{\begin{centering}
\includegraphics[scale=0.5]{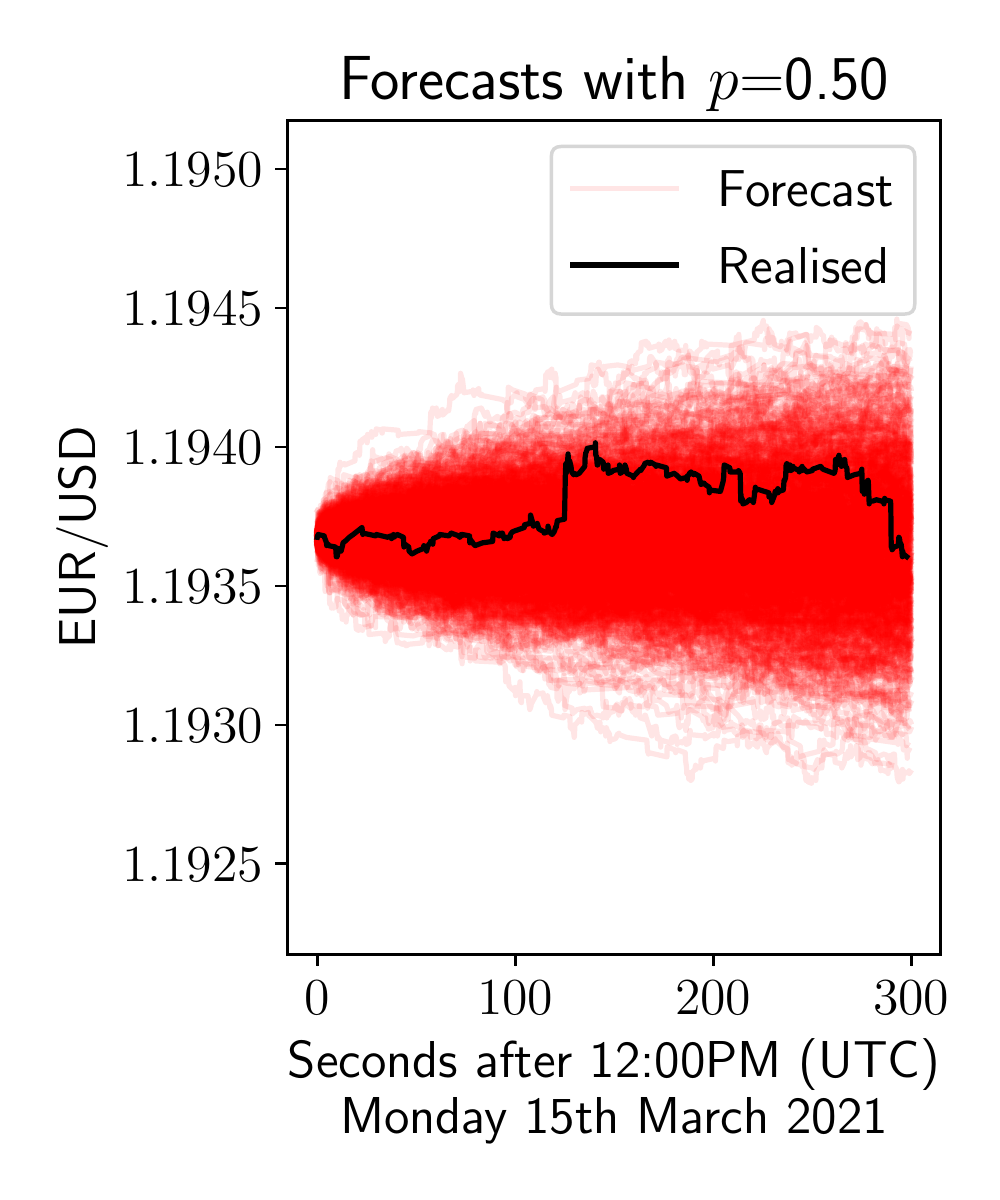}
\par\end{centering}
}\subfloat[Forecasts \& historical data $p=\frac{3}{4}$]{\begin{centering}
\includegraphics[scale=0.5]{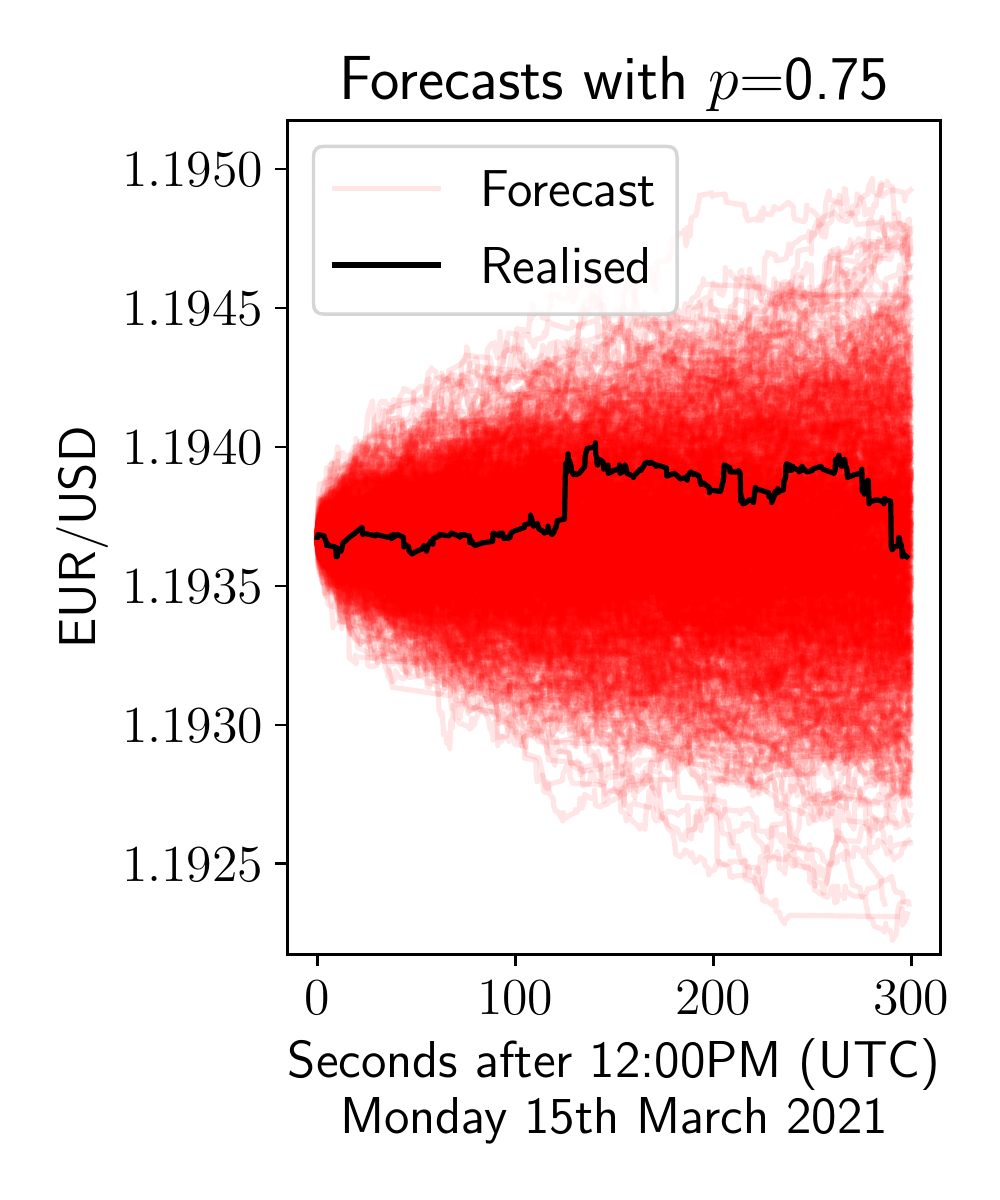}
\par\end{centering}
} \end{adjustbox} 
\par\end{centering}
\centering{}\caption{Forecast plots}
\label{fig:forecastseurusd}
\end{figure}

Initially, we plot forecasts from the model in \ref{fig:forecastseurusd}
where we plot realisations from the model for 5 minutes after midday
on 15th March 2021. We do this for three values of $p$ and using
the previous week to estimate the empirical distribution in (\ref{eq:empdists}).
We can see that the variance is increasing as $p$ increases, consistent
with the continuous time random walk model. 

\begin{figure}[H]
 \begin{adjustbox}{center}\subfloat[Measure of forecast quality (lower the better) as a function of $p$]{\begin{centering}
\includegraphics[scale=0.5]{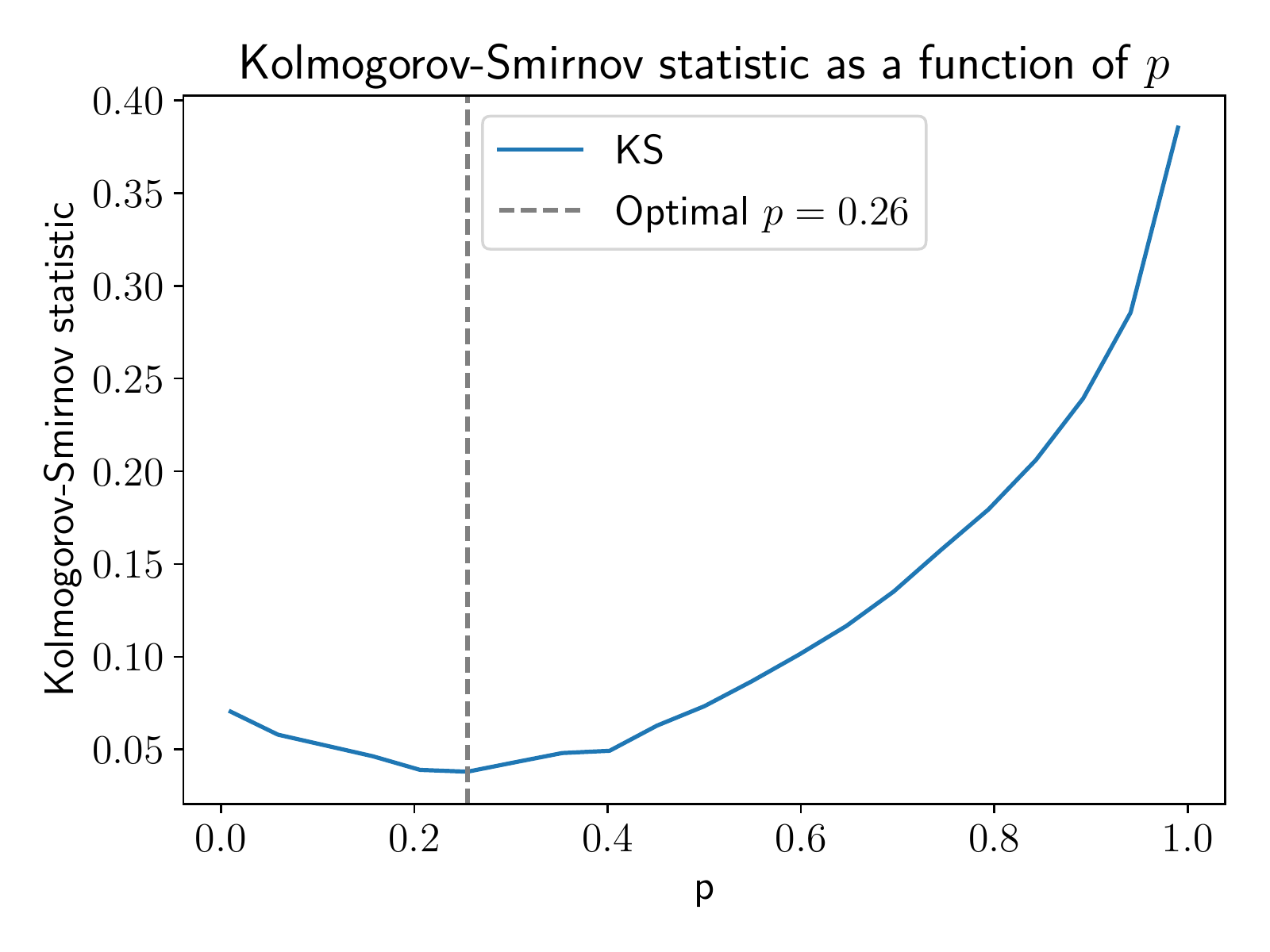}
\par\end{centering}
\label{fig:ksvaryp}}\subfloat[Variance for EUR/USD as a function of $p$]{\begin{centering}
\includegraphics[scale=0.5]{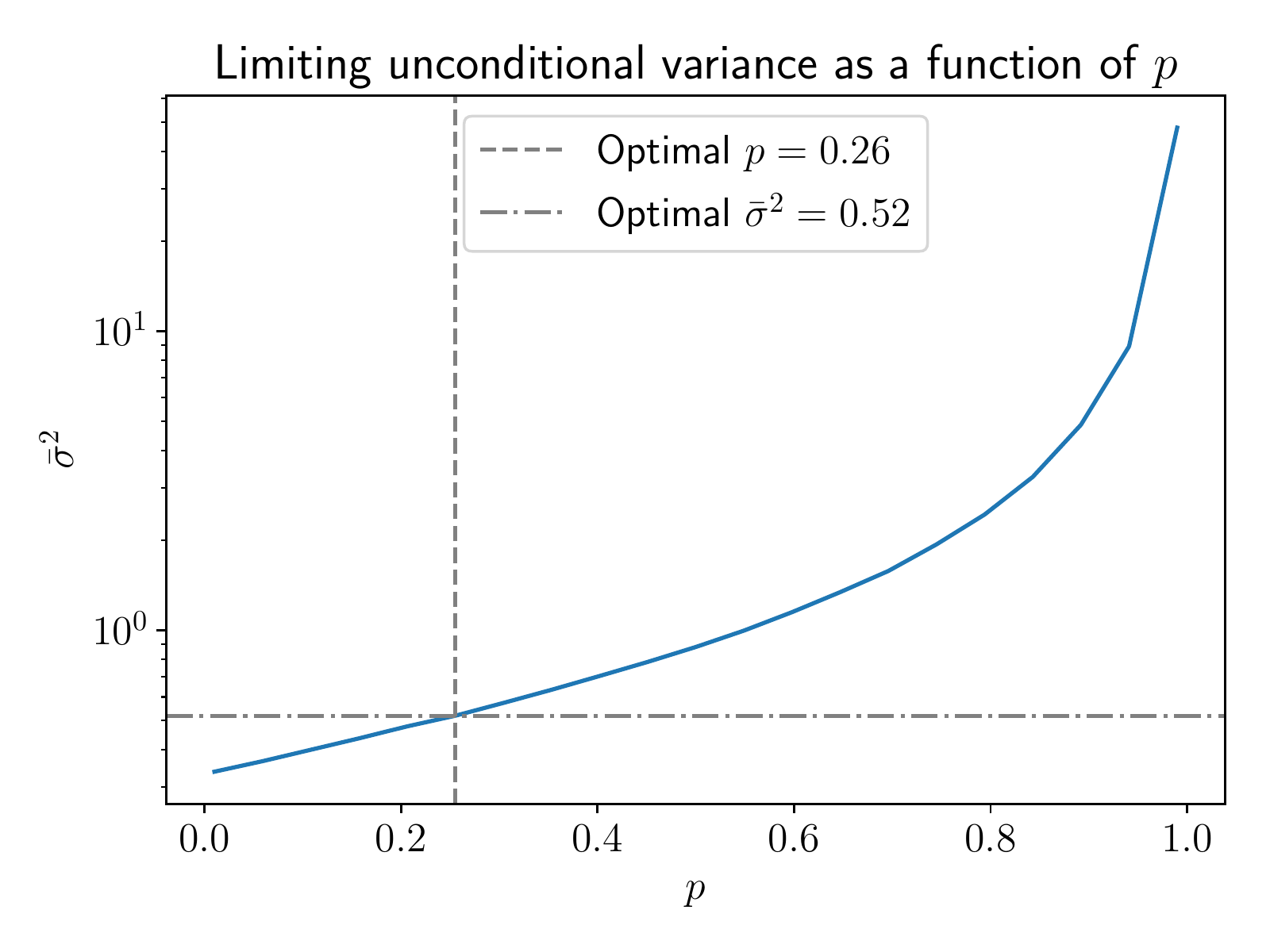}
\par\end{centering}
\label{fig:varvaryingpeurusd}} \end{adjustbox} 
\centering{}\caption{Fitting the model to a week of EUR/USD data and predicting on the
subsequent week.}
\end{figure}

Subsequently, we assess the performance of the model at forecasting
through a simulation based approach in the spirit of \citep{andersen2003modeling}.
For a given value of $p$, such an approach iteratively performs Monte
Carlo simulations to obtain the distribution of price process at a
given time point. Let $\{\hat{Z}_{t}^{(i)}\}$ represent the predicted
distribution of the process $Z_{t}$ at time $t$. As in \citep{andersen2003modeling},
we are then able to construct a statistic through taking the probability
integral transform of $Z_{t}$ through the empirical distribution
function $\hat{F}(\{\hat{Z}_{t}^{(i)}\})$, which will be approximately
uniformly distributed if the predictions are accurate. For varying
values of $p$, we assess the quality of forecast predictions using
the Kolmogorov-Smirnov distance of the probabilty integral transforms
from the uniform distribution. Forecasts were made every 5 minutes
over periods of 5 minutes from Monday 0:00AM (UTC) 15th March 2021
to Saturday 0:00AM (UTC) 20th March 2021. Figure \ref{fig:ksvaryp}
shows how the Kolmogorov-Smirnov distance (our measure of prediction
accuracy) varies as $p$ varies, using 100,000 Monte Carlo simulations.
We see that there is a clear minimum at $p=0.26.$ We cross-reference
this with the variance predicted by the model through simulations
at varying levels of $p$, shown in Figure \ref{fig:varvaryingpeurusd},
where we see that the variance is less than the variance in the absence
of these correlations ($p=0.5$), to the extent that it is deflated
to approximately 52\% of the variance without correlations. Indeed
we see that over this week the prevailing trading style is most likely
therefore mean-reverting. 

\section{Discussion and conclusion}

We have explored using a probabilistic model for investment styles
to show that the variance of a financial asset is directly dependent
on the probability of moving in the same direction on successive days.
The theoretical analysis shows that variance may actually be reduced
through reversal strategies - capturing the case that the asset is
more likely to move in opposing directions on subsequent days. We
have applied a simple model to US stock data, showing that such a
regime is indeed prevalent in 97 of the largest stocks and thereby
proven that relative to a random walk the variance of these stocks
is actually \emph{reduced} as a result of this frenetic behaviour.
Indeed such a result suggests that these stocks are actually \emph{more
}predictable than a random walk due to this artefacct. A similar result
is also shown for higher-frequency data in the case of the EUR/USD
to a more exagerated extent. Similar ideas have been explored from
a different perspective with the identification of a Hurst exponent
of a stochastic process, quantifying the asymptotic behaviour of the
range of a stochastic processs \citep{qian2004hurst}, though in general
describe the behaviour only qualitatively. 

The models considered above allow for many more flexible generalisations.
In particular, \citep{lebaron1992some} remarks that period of high
volatility are typically occur during periods of low autocorrelation
(and vice verser). It would be interesting to allow for variable volatility
in the above model, exploring dynamically changing correlation parameters
$p$. Alternatively, it would also be interesting to consider the
sign process for multiple assets simultanesously rather than focussing
- as is the case here - on a single asset. Similarly, it may be fruiful
to consider a sign process where subsequent values depend on more
than one lag of the process, thereby providing a still more flexible
model. Finally, a principled incorporation of processes exhibiting
long-term drift would also provide additional insight into asset price
behaviour. 

More generally, that autocorrelation contributes to larger variance
is well known within the computational statistics literature. In particular,
the autocorrelation in a class of algorithms referred to as Markov
chain Monte Carlo algorithms typically serves to increase the variance
arising from the Monte Carlo noise \citep{robert2013monte} in a directly
analagous fashion to the increase in variance as defined above.

\pagebreak{}

\bibliographystyle{plainnat}
\bibliography{passivebib}

\pagebreak{}

\appendix

\section{Appendix}

\subsection{Proof of Proposition \ref{prop:semipprop}}

\label{subsec:proofprop}

We analyse first of all, the increment process defined as 
\[
Z_{n}=\sum_{i=1}^{n}S_{i}\delta_{i}
\]
$S_{i}\in\{-1,1\}$. By symmetry we have that $\mathds{E}[Z_{n}]=0$,
so that the variance is given as 
\begin{align}
\mathds{V}[Y_{n}] & =\mathds{E}\left[\left(\sum_{i=1}^{n}S_{i}\delta_{i}\right)^{2}\right]\label{eq:varY}\\
 & =\mathds{E}\left[\sum_{i=1}^{n}\left(\left(S_{i}\delta_{i}\right)^{2}+2\sum_{j=1}^{i-1}\delta_{i}S_{i}\delta_{j}S_{j}\right)\right]\\
 & .=\mathds{E}_{\nu}[\delta^{2}]\sum_{i=1}^{n}\mathds{E}[S_{i}^{2}]+2\mathds{E}_{\nu}[\delta]^{2}\sum_{i=1}^{n}\sum_{j=1}^{i-1}\mathds{E}[S_{i}S_{j}]
\end{align}
where the final line follows from independence of the magnitudes $\delta_{i}$. 

We state the $n$-step transition matrix of $S_{n}$ as 
\[
P^{n}=\frac{1}{2}\left(\begin{array}{cc}
1+\delta^{n} & 1-\delta^{n}\\
1-\delta^{n} & 1+\delta^{n}
\end{array}\right)
\]
where $[P^{n}]_{ab}=\mathds{P}[S_{n}=2a-1|S_{1}=2b-1]$ and $\delta:=2p-1$
for $(a,b)\in\{0,1\}^{2}$. We see this by induction in that
\[
P^{1}=\left(\begin{array}{cc}
p & 1-p\\
1-p & p
\end{array}\right)=\frac{1}{2}\left(\begin{array}{cc}
1+\delta & 1-\delta\\
1-\delta & 1+\delta
\end{array}\right)
\]
and
\begin{align*}
P^{n}P & =\frac{1}{2}\left(\begin{array}{cc}
1+\delta^{n} & 1-\delta^{n}\\
1-\delta^{n} & 1+\delta^{n}
\end{array}\right)\frac{1}{2}\left(\begin{array}{cc}
1+\delta & 1-\delta\\
1-\delta & 1+\delta
\end{array}\right)\\
 & =\frac{1}{4}\left(\begin{array}{cc}
(1+\delta)(1+\delta^{n})+(1-\delta)(1-\delta^{n}) & (1-\delta)(1+\delta^{n})+(1+\delta)(1-\delta^{n})\\
(1+\delta)(1-\delta^{n})+(1-\delta)(1+\delta^{n}) & (1-\delta)(1-\delta^{n})+(1+\delta)(1+\delta^{n})
\end{array}\right)\\
 & =\frac{1}{4}\left(\begin{array}{cc}
2+2\delta^{n+1} & 2-2\delta^{n+1}\\
2-2\delta^{n+1} & 2+2\delta^{n+1}
\end{array}\right)=P^{n+1}
\end{align*}
We have that $\mathds{E}[S_{i}^{2}]=1$ and by symmetry we see that
\begin{align*}
\mathds{E}[S_{i}S_{j}] & =\mathds{E}\left[S_{i}\mathds{E}[S_{j}|S_{i}]\right]\\
 & =2\left(\frac{1+\delta^{j-i}}{2}-\frac{1-\delta^{j-i}}{2}\right)\frac{1}{2}=(2p-1)^{j-i}
\end{align*}
so that from Equation (\ref{eq:varY}) we have 
\begin{align}
\mathds{V}[Y_{n}] & =n\mathds{E}_{\nu}[\delta^{2}]+2\mathds{E}_{\nu}[\delta]^{2}\sum_{i=1}^{n}\sum_{j=1}^{i-1}(2p-1)^{j-i}\nonumber \\
 & =n\mathds{E}_{\nu}[\delta^{2}]+\mathds{E}_{\nu}[\delta]^{2}\sum_{i=1}^{n}\left(\frac{(2p-1)-(2p-1)^{i}}{1-p}\right)\label{eq:varasympexp}
\end{align}
Which, after rescaling, we see that 
\begin{align*}
\lim_{n\rightarrow\infty}\frac{1}{n}\mathds{V}[Y_{n}] & =\mathds{E}_{\nu}[\delta^{2}]+\mathds{E}_{\nu}[\delta]^{2}\frac{2p-1}{1-p}-\lim_{n\rightarrow\infty}\frac{1}{n}\sum_{i=1}^{n}\frac{(2p-1)^{i}}{1-p}\\
 & =\mathds{E}_{\nu}[\delta^{2}]+\mathds{E}_{\nu}[\delta]^{2}\left(\frac{2p-1}{1-p}\right)
\end{align*}
where the final term tends to 0 as $x^{n}/n\rightarrow0$ for $|x|\le1$. 

\subsection{Analysis of a high-frequency model}

\label{subsec:continuoustimeanalysis}

We define the times at which the direction changes as a renewal process,
i.e. we assume the time between successive events is itself a random
variable, which we denote by $\tau_{1},\tau_{2},...$. We define a
counting process as $N(t)=\sup\{n:\sum_{i=1}^{n}\tau_{i}\le t\}$.
Considering the definition of the Markov sign process $(S_{n})$ and
magnitdue increments $(\delta_{n})$ as before, the resulting process
is then expressed as 
\[
Z_{t}=\sum_{i=1}^{N(t)}S_{i}\delta_{i}.
\]
The analysis of a daily model, can be seen as a special case where
for all $i$ , $\tau_{i}$ is deterministically set equal to 1.

We are interested in how the parameter $p$ affects the long-run variance
of the resulting stochastic process. As a result, we define the limiting
variance as 
\[
\bar{\sigma}^{2}:=\lim_{t\rightarrow\infty}\mathds{V}\left[\frac{1}{\sqrt{t}}Z_{t}\right]
\]
Following a similar analysis as for Proposition \ref{prop:semipprop},
we are able to provide an analagous expression for the limiting variance
as follows.
\begin{thm}
\label{thm:mainres}Under the same assumptions as Proposition \ref{prop:semipprop},
then assuming $(\tau_{i})_{i\ge1}$ are independent of $(S_{i})_{i\ge1},(\delta_{i})_{i\ge1}$
and $\mu_{\tau}:=\lim_{t\rightarrow\infty}\mathds{E}_{N}\left[\frac{N(t)}{t}\right]<\infty$,
we have that the limiting variance is 
\begin{equation}
\bar{\sigma}^{2}=\mu_{\tau}\left(\mathds{E}_{\nu}[\delta^{2}]+\mathds{E}_{\nu}[\delta]^{2}\left(\frac{2p-1}{1-p}\right)\right)\label{eq:continuousvar}
\end{equation}
\end{thm}

We allow for quite a general structure on the arrival time process,
noting they do not necessarily exhibit finite variance and are not
necessarily i.i.d. (they could themselves be a Markov process for
example). From Theorem \ref{thm:mainres}, we see (as before) that
Equation (\ref{eq:continuousvar}) can be simplified in the special
case of a deterministic magnitude distribution, suggesting

\[
\bar{\sigma}^{2}\propto\frac{p}{1-p}
\]
providing a parallel with the discrete-time process.

\label{subsec:proofmainres}
\begin{proof}
Considering now the Markov renewal process $Z_{t}$, and defining
$\mu_{\tau}:=\lim_{t\rightarrow\infty}\mathds{E}_{N}\left[\frac{N(t)}{t}\right]$,
we are able to condition on $N(t)=n$ and take expectations so that 

\begin{align}
\mathds{V}\left[\frac{1}{\sqrt{t}}Z_{t}\right] & =\mathds{E}\left[\left(\frac{1}{\sqrt{t}}\sum_{i=1}^{N(t)}W_{i}\right)^{2}\right]\nonumber \\
 & =\mathds{E}_{N}\left[\mathds{E}\left[\frac{1}{t}\left(\sum_{i=1}^{n}W_{i}\right)^{2}|N(t)=n\right]\right]\nonumber \\
 & =\mathds{E}_{N}\left[\frac{N(t)}{t}\mathds{E}_{\nu}[\delta^{2}]+\frac{N(t)}{t}\mathds{E}_{\nu}[\delta]^{2}\frac{2p-1}{1-p}-\frac{1}{t}\mathds{E}_{\nu}[\delta]^{2}\sum_{i=1}^{N(t)}\frac{(2p-1)^{i}}{1-p}\right]\nonumber \\
 & =\mathds{E}_{N}\left[\frac{N(t)}{t}\right]\left(\mathds{E}_{\nu}[\delta^{2}]+\mathds{E}_{\nu}[\delta]^{2}\left(\frac{2p-1}{1-p}\right)\right)-\mathds{E}_{\nu}[\delta]^{2}\mathds{E}_{N}\left[\frac{1}{(1-p)t}\sum_{i=1}^{N(t)}(2p-1)^{i}\right]\nonumber \\
 & =\mathds{E}_{N}\left[\frac{N(t)}{t}\right]\left(\mathds{E}_{\nu}[\delta^{2}]+\mathds{E}_{\nu}[\delta]^{2}\left(\frac{2p-1}{1-p}\right)\right)+\mathds{E}_{\nu}[\delta]^{2}\mathds{E}_{N}\left[\frac{1}{t}\frac{(1-2p)}{(1-p)2p}\left(1-(2p-1)^{N(t)}\right)\right]\label{eq:lastline}
\end{align}
We have by Jensen's inequality and the triangle inequality
\begin{align*}
\left|\mathds{E}_{N}\left[\frac{1}{t}\frac{(1-2p)}{(1-p)2p}\left(1-(2p-1)^{N(t)}\right)\right]\right| & \le\frac{1}{t}\left|\frac{(1-2p)}{(1-p)2p}\right|\mathds{E}_{N}\left[\left|\left(1-(2p-1)^{N(t)}\right)\right|\right]|\\
 & \le\frac{1}{t}\left|\frac{(1-2p)}{(1-p)2p}\right|\mathds{E}_{N}\left[1+\left|(2p-1)^{N(t)}\right|\right]\\
 & \le\frac{1}{t}\left|\frac{(1-2p)}{(1-p)p}\right|
\end{align*}
as $|(2p-1)^{N(t)}|\le1$ we have that. As such, continuing from equation
(\ref{eq:lastline}) we have the limiting variance given by 
\[
\lim_{t\rightarrow\infty}\mathds{V}\left[\frac{1}{\sqrt{t}}Z_{t}\right]=\mu_{\tau}\left(\mathds{E}_{\nu}[\delta^{2}]+\mathds{E}_{\nu}[\delta]^{2}\left(\frac{2p-1}{1-p}\right)\right).
\]
\end{proof}

\subsection{US daily stock model validation}

\label{subsec:USmodelvalidation}

We here describe the model validation used to justify the model proposed
in Equation (\ref{eq:semipflex}) for US daily stock data. In particular
we check
\begin{enumerate}
\item The empirical distribution of $\delta_{i}$ is symmetric
\item The observed $\delta_{i}$ are uncorrelated with $S_{i}$
\end{enumerate}
Strictly speaking, we should investigate independence between $(\delta_{i})$
and $(S_{i})$ though test for correlation for convenience. To test
whether the empirical distribution of $\delta_{i}$ is symmetric,
we employ the two-sample Kolmogorov-Smirnov test on increments of
the log-price process that are non-negative against those that are
non-positive. Of the 97 tests performed 5.15\% of those were significant
at 5\%, which is perhaps a little high, though within acceptable tolerance.
To test the correlation between between $(\delta_{i})$ and $(S_{i})$
we estimate the $p$-value corresponding to Pearson's correlation.
In this case, of the 97 tests performed 2.06\% were significant at
5\% thereby not providing evidence to reject the null that the two
samples are uncorrelated.

\subsection{US daily stock data}

\label{subsec:usstocktables}

Summary data of the 100 large market capitalisation US stocks is provided
in Table \ref{tab:US1} and \ref{tab:US2}.

\begin{table}
 \begin{adjustbox}{center}

\begin{tabular}{|c|c|c|c|c|}
\hline 
Symbol & Name & Price (intraday) & Volume & Market cap (\$)\tabularnewline
\hline 
\hline 
AAPL & Apple Inc. & 146.09 & 48.802M & 2.415T\tabularnewline
\hline 
MSFT & Microsoft Corporation & 288.33 & 13.927M & 2.167T\tabularnewline
\hline 
GOOGL & Alphabet Inc. & 2,738.26 & 857,805 & 1.834T\tabularnewline
\hline 
AMZN & Amazon.com, Inc. & 3,341.87 & 2.035M & 1.692T\tabularnewline
\hline 
FB & Facebook, Inc. & 361.61 & 7.543M & 1.02T\tabularnewline
\hline 
TSLA & Tesla, Inc. & 713.76 & 14.543M & 706.633B\tabularnewline
\hline 
BRK-B & Berkshire Hathaway Inc. & 287.23 & 3.67M & 656.99B\tabularnewline
\hline 
TSM & Taiwan Semiconductor Manufacturing Company Limited & 118.22 & 4.007M & 613.089B\tabularnewline
\hline 
TCEHY & Tencent Holdings Limited & 60.55 & 7.093M & 587.49B\tabularnewline
\hline 
BABA & Alibaba Group Holding Limited & 195.25 & 14.287M & 537.918B\tabularnewline
\hline 
JPM & JPMorgan Chase \& Co. & 157.33 & 8.961M & 470.127B\tabularnewline
\hline 
V & Visa Inc. & 240.00 & 5.161M & 526.654B\tabularnewline
\hline 
NVDA & NVIDIA Corporation & 202.95 & 13.289M & 505.751B\tabularnewline
\hline 
JNJ & Johnson \& Johnson & 173.71 & 3.801M & 457.288B\tabularnewline
\hline 
WMT & Walmart Inc. & 145.58 & 5.259M & 407.937B\tabularnewline
\hline 
BAC & Bank of America Corporation & 40.67 & 48.111M & 342.234B\tabularnewline
\hline 
UNH & UnitedHealth Group Incorporated & 410.87 & 1.466M & 387.416B\tabularnewline
\hline 
MA & Mastercard Incorporated & 370.68 & 2.362M & 365.778B\tabularnewline
\hline 
HD & The Home Depot, Inc. & 328.76 & 1.876M & 349.557B\tabularnewline
\hline 
PG & The Procter \& Gamble Company & 142.18 & 4.352M & 345.455B\tabularnewline
\hline 
RHHBY & Roche Holding AG & 49.07 & 1.145M & 342.978B\tabularnewline
\hline 
NSRGY & Nestlé S.A. & 123.50 & 131,247 & 342.528B\tabularnewline
\hline 
PYPL & PayPal Holdings, Inc. & 278.15 & 3.706M & 326.835B\tabularnewline
\hline 
ASML & ASML Holding N.V. & 788.68 & 472,372 & 325.107B\tabularnewline
\hline 
DIS & The Walt Disney Company & 176.66 & 4.926M & 320.979B\tabularnewline
\hline 
ADBE & Adobe Inc. & 629.22 & 996,552 & 299.76B\tabularnewline
\hline 
NKE & NIKE, Inc. & 171.77 & 3.396M & 271.706B\tabularnewline
\hline 
CMCSA & Comcast Corporation & 58.28 & 12.086M & 267.49B\tabularnewline
\hline 
PFE & Pfizer Inc. & 45.98 & 30.369M & 257.382B\tabularnewline
\hline 
LLY & Eli Lilly and Company & 267.16 & 2.791M & 255.56B\tabularnewline
\hline 
TM & Toyota Motor Corporation & 180.57 & 155,646 & 252.233B\tabularnewline
\hline 
ORCL & Oracle Corporation & 89.90 & 5.678M & 251.001B\tabularnewline
\hline 
KO & The Coca-Cola Company & 56.65 & 8.185M & 244.537B\tabularnewline
\hline 
CRM & salesforce.com, inc. & 249.32 & 3.298M & 242.222B\tabularnewline
\hline 
XOM & Exxon Mobil Corporation & 57.20 & 15.209M & 242.16B\tabularnewline
\hline 
CSCO & Cisco Systems, Inc. & 55.47 & 6.963M & 233.762B\tabularnewline
\hline 
NVO & Novo Nordisk A/S & 100.65 & 1.031M & 231.352B\tabularnewline
\hline 
NFLX & Netflix, Inc. & 519.97 & 1.302M & 230.137B\tabularnewline
\hline 
VZ & Verizon Communications Inc. & 55.12 & 12.064M & 228.203B\tabularnewline
\hline 
DHR & Danaher Corporation & 307.97 & 1.368M & 219.86B\tabularnewline
\hline 
INTC & Intel Corporation & 54.04 & 14.204M & 219.24B\tabularnewline
\hline 
ABT & Abbott Laboratories & 123.16 & 4.265M & 218.341B\tabularnewline
\hline 
WFC-PO & Wells Fargo \& Company & 25.27 & 120,697 & 213.443B\tabularnewline
\hline 
PEP & PepsiCo, Inc. & 154.35 & 2.188M & 213.329B\tabularnewline
\hline 
TMO & Thermo Fisher Scientific Inc. & 541.16 & 831,562 & 212.691B\tabularnewline
\hline 
NVS & Novartis AG & 91.81 & 1.226M & 207.074B\tabularnewline
\hline 
ACN & Accenture plc & 319.52 & 1.205M & 202.619B\tabularnewline
\hline 
ABBV & AbbVie Inc. & 114.06 & 5.201M & 201.565B\tabularnewline
\hline 
\end{tabular}

 \end{adjustbox}\caption{100 largest US stocks by market capitalisation 2021-08-09}
\label{tab:US1}
\end{table}
\begin{table}
 \begin{adjustbox}{center}

\begin{tabular}{|c|c|c|c|c|}
\hline 
Symbol & Name & Price (intraday) & Volume & Market cap (\$)\tabularnewline
\hline 
\hline 
WFC & Wells Fargo \& Company & 48.65 & 26.408M & 199.777B\tabularnewline
\hline 
AVGO & Broadcom Inc. & 484.68 & 569,893 & 198.845B\tabularnewline
\hline 
VWDRY & Vestas Wind Systems A/S & 13.18 & 241,291 & 198.801B\tabularnewline
\hline 
T & AT\&T Inc. & 27.85 & 21.162M & 198.849B\tabularnewline
\hline 
MRNA & Moderna, Inc. & 484.47 & 41.879M & 195.554B\tabularnewline
\hline 
COST & Costco Wholesale Corporation & 440.47 & 1.244M & 194.718B\tabularnewline
\hline 
BHP & BHP Billiton Limited & 76.54 & 979,561 & 194.584B\tabularnewline
\hline 
SHOP & Shopify Inc. & 1,549.99 & 1.083M & 194.256B\tabularnewline
\hline 
CVX & Chevron Corporation & 100.25 & 8.423M & 193.874B\tabularnewline
\hline 
MRK & Merck \& Co., Inc. & 75.32 & 7.175M & 190.715B\tabularnewline
\hline 
SBRCY & Sberbank of Russia & 17.70 & 138,498 & 184.988B\tabularnewline
\hline 
CICHY & China Construction Bank Corporation & 14.17 & 194,040 & 180.599B\tabularnewline
\hline 
C & Citigroup Inc. & 71.52 & 14.185M & 144.956B\tabularnewline
\hline 
TMUS & T-Mobile US, Inc. & 142.99 & 3.012M & 178.447B\tabularnewline
\hline 
MPNGY & Meituan & 57.56 & 111,865 & 176.389B\tabularnewline
\hline 
MS & Morgan Stanley & 100.74 & 8.091M & 183.806B\tabularnewline
\hline 
TXN & Texas Instruments Incorporated & 190.45 & 2.315M & 175.825B\tabularnewline
\hline 
AZN & AstraZeneca PLC & 56.36 & 9.63M & 175.727B\tabularnewline
\hline 
MCD & McDonald's Corporation & 234.68 & 1.858M & 175.259B\tabularnewline
\hline 
SAP & SAP SE & 146.56 & 463,896 & 174.447B\tabularnewline
\hline 
VWAGY & Volkswagen AG & 34.79 & 437,040 & 174.401B\tabularnewline
\hline 
MDT & Medtronic plc & 126.92 & 2.949M & 170.568B\tabularnewline
\hline 
UPS & United Parcel Service, Inc. & 190.98 & 1.752M & 166.254B\tabularnewline
\hline 
QCOM & QUALCOMM Incorporated & 146.92 & 4.536M & 165.726B\tabularnewline
\hline 
BBL & BHP Group & 63.80 & 1.161M & 161.725B\tabularnewline
\hline 
PNGAY & Ping An Insurance (Group) Company of China, Ltd. & 17.68 & 502,175 & 161.598B\tabularnewline
\hline 
SE & Sea Limited & 307.14 & 1.744M & 161.075B\tabularnewline
\hline 
RDS-A & Royal Dutch Shell plc & 41.04 & 4.047M & 160.118B\tabularnewline
\hline 
RDS-B & Royal Dutch Shell plc & 40.35 & 4.187M & 159.578B\tabularnewline
\hline 
NEE & NextEra Energy, Inc. & 80.56 & 4.83M & 158.039B\tabularnewline
\hline 
HON & Honeywell International Inc. & 228.23 & 1.306M & 157.57B\tabularnewline
\hline 
LIN & Linde plc & 303.26 & 900,463 & 156.607B\tabularnewline
\hline 
PM & Philip Morris International Inc. & 99.20 & 1.609M & 154.607B\tabularnewline
\hline 
GS & The Goldman Sachs Group, Inc. & 399.88 & 3.296M & 134.798B\tabularnewline
\hline 
BMY & Bristol-Myers Squibb Company & 67.38 & 6.647M & 149.726B\tabularnewline
\hline 
UL & Unilever PLC & 57.26 & 1.25M & 149.476B\tabularnewline
\hline 
INTU & Intuit Inc. & 535.23 & 724,449 & 146.256B\tabularnewline
\hline 
AAGIY & AIA Group Limited & 47.88 & 192,823 & 145.212B\tabularnewline
\hline 
RY & Royal Bank of Canada & 102.80 & 562,572 & 147.145B\tabularnewline
\hline 
UNP & Union Pacific Corporation & 219.95 & 1.615M & 143.434B\tabularnewline
\hline 
PROSY & Prosus N.V. & 17.52 & 1.026M & 142.991B\tabularnewline
\hline 
CHTR & Charter Communications, Inc. & 765.50 & 541,813 & 140.716B\tabularnewline
\hline 
SBUX & Starbucks Corporation & 117.94 & 4.439M & 139.063B\tabularnewline
\hline 
RIO & Rio Tinto plc & 85.57 & 2.071M & 138.546B\tabularnewline
\hline 
SCHW & The Charles Schwab Corporation & 73.42 & 6.079M & 138.496B\tabularnewline
\hline 
HDB & HDFC Bank Limited & 75.16 & 629,583 & 138.463B\tabularnewline
\hline 
BLK & BlackRock, Inc. & 901.97 & 276,240 & 137.369B\tabularnewline
\hline 
BA & The Boeing Company & 232.27 & 8.298M & 136.146B\tabularnewline
\hline 
AXP & American Express Company & 170.78 & 2.118M & 135.673B\tabularnewline
\hline 
\end{tabular}

 \end{adjustbox}

\caption{Largest US stocks by market capitalisation 2021-08-09}
\label{tab:US2}
\end{table}

\end{document}